\documentclass[aps,pre,twocolumn,showpacs,nofootinbib]{revtex4-1}
\usepackage[dvips]{graphicx}
\usepackage{bm,dcolumn}
\usepackage{mathrsfs}
\usepackage{lipsum}
\usepackage{graphicx,epsfig}
\usepackage{psfrag}
\usepackage{bm}
\usepackage{epstopdf}
\usepackage{latexsym}
\usepackage{multirow}
\usepackage{amsmath,amssymb,amsfonts}
\usepackage{enumerate}
\usepackage[utf8]{inputenc}
\usepackage{color,soul}
\usepackage{float}
\usepackage{tikz}
\allowdisplaybreaks
\begin{document}
\date{\today}

\title{ Role of curvature and domain shape on Turing patterns
}
\author{Sankaran Nampoothiri} \email[email:] {sankaran@iisertvm.ac.in}
\author{Amal Medhi} \email{amedhi@iisertvm.ac.in}
\affiliation{ School of Physics, Indian Institute of Science Education and Research Thiruvananthapuram, India}
\begin{abstract}
We consider pattern formation using reaction-diffusion equation on various
non-uniformly curved surfaces. We explore how, in general, curvature
and, in particular the domain shape would affect the pattern formation in 
these geometries. 
As examples, we study stripe and spot 
patterns on a torus, and on an ellipsoid. Our results show that the 
curvature and domain shape can control the orientation of stripe pattern as well as 
the size and number of spots. Our
results also indicate that by controlling the curvature and shape, one can
drive the chemicals to a preferred region. Specifically on a torus, 
curvature and shape can guide the chemicals more on to outer side than inner side. This
result may prove important in the studies of self-organization of molecules
in biological membranes.
\end{abstract}

\maketitle
\section{INTRODUCTION}
\paragraph*{}
 Pattern formation is an ubiquitous phenomenon in nature. Hence its modeling is of fundamental
 importance in many fields. After the seminal work
 of Turing~\cite{turing}, reaction-diffusion systems
 are widely used for mathematical modeling of spatial pattern formation~\cite{murray,biology,vegetation1,vegetation2}.
 In reaction-diffusion
 (RD), chemicals are allowed to react and diffuse so as to produce 
 a patterned steady state. 
  RD equations are routinely used in modeling
 the skin patterns in fish~\cite{shoji}, mammals~\cite{bard}, snakes~\cite{snake},
 and leopards~\cite{maini}.
 Recently 
 RD equations
 are also used in understanding the spatial organization of molecules
 in biological membranes~\cite{protein}. For
example, RD models are proposed to explain oscillations
of Min protein system in E.coli cells~\cite{howard,min1,min2}.
\paragraph*{}
In most studies, RD equations are usually considered on flat geometries
for studying patterns. 
Pattern formation on animal surfaces is an example  where 
 the surface is not flat. Another example is the spatial organization of
 molecules on biological membranes where
  spherical, saddle and toroidal shaped membranes are
 common in nature~\cite{membranecurvature,curvature}. It 
 is reasonable to assume that curvature and shape of the 
 surface can play an important role in the formation of spatial patterns. 
 For instance, a recent study shows that 
   membrane geometry can control spatial organization of molecules
 in biological membranes~\cite{curvature}. 
 The geometry can also be responsible for some of the complex patterns
observed on animal surfaces~\cite{murray,venkataraman}. 
\paragraph*{}
Owing to the importance of geometry on pattern formation,
some of the previous studies have considered RD equations 
 on uniformly curved surfaces. For instance, Varea et al.~\cite{varea} analyzed
 a generic RD system on a spherical surface, and Zykov et al.~\cite{zykov} studied evolution of 
 spiral waves on the 
 same surface. Meandering of spiral 
 waves on a sphere is thoroughly analyzed using RD equation~\cite{meandering}. RD equation 
 on a hemisphere
 is used to model the spot formation on the hard wings of lady beetles~\cite{ladybeetle}.
 \paragraph*{}
 Recently, some studies have been initiated to understand
 RD equations on non-uniformly curved surfaces. To some extent, the role 
 of growth and curvature in RD is attempted in the work of 
 Plaza et al.~\cite{plaza}. The
 process of parr-marks formation on fish skins are studied where shape of the skin is 
 modeled as growing elliptic cylinder~\cite{venkataraman}. 
 These studies suggest that 
 the curvature and domain shape can strongly influence the formation of spatial patterns 
 on non-flat geometries.  More recently, study on nucleation of RD waves 
 on curved surfaces~\cite{nucleation} and spiral waves 
 on curved surfaces~\cite{new1} again suggest the 
 importance of curvature and domain shape in RD studies.
 \paragraph*{}
 In this work, we aim to study pattern formation using RD equation
 on various non-uniformly curved surfaces. In particular, we explore the role of
 curvature and domain shape
  in determining the size and
 the distribution of spot patterns, and orientation
 of stripe patterns on curved surfaces. We consider RD equation, say, on a torus for 
 different $R$ and $r$ values, where $R$ is major radius and $r$ is minor radius but
 keeping the area constant. We then analyze how 
 the number as well as the size and position of the spots and the orientation
 of stripe patterns vary as we change $R$ and $r$? 
  Similarly, 
 we also extend our
 analysis to ellipsoid surface by varying the shape.
 \paragraph*{}
 The paper is organized as follows. In Sec.~II, we give a brief description
 about the model used for 
 the study. In Sec.~III, we study patterns on various 
 non-uniformly curved surfaces, and analyze
  the role of curvature and domain shape in the formation of patterns on these surfaces.
 Here we obtain spot and stripe patterns on a torus, and on an ellipsoid. Finally
 we conclude our results in Sec.~IV.
  \section{model}
 The dynamics of RD system on a given curved surface is governed
 by following set of equations
 \begin{subequations}
\begin{eqnarray}
 \frac{\partial U}{\partial t}&=F_{1}(U,V)+D_{U}\bigtriangleup_{LB} u ,\\                                                      
 \frac{\partial V}{\partial t}&=F_{2}(U,V)+D_{V}\bigtriangleup_{LB} v,
 \label{eq:rd_equation}
\end{eqnarray}
\end{subequations}
 where $U$, $V$ are the concentrations of chemicals,
 $F_{1}(U,V),F_{2}(U,V)$
 represents the reaction kinetics, and $D_{U},D_{V}$ are diffusion coefficients
 of the chemicals $U$ and $V$ respectively, and
 $\bigtriangleup_{LB}$ is the Laplace-Beltrami operator on curved surface.
 \paragraph*{}
 Following Barrio et al.~\cite{barrio} we consider specific form of RD equations
 \begin{eqnarray}
 \begin{aligned}
 \frac{\partial u}{\partial t}&=\alpha u(1-r_{1}v^{2})+v(1-r_{2}u)+D\delta \bigtriangleup_{LB}u,\\
 \frac{\partial v}{\partial t}&=\beta v(1+\frac{\alpha r_{1}}{\beta}uv)+u(\gamma+r_{2}v)+
 \delta\bigtriangleup_{LB} v,
 \end{aligned}
 \label{eq:rd_barrio}
\end{eqnarray}
 where $u$ and $v$ are small deviations of U and V from homogeneous steady state
 values. The parameters
 $\alpha$, $\beta$, $\gamma$ are related to production and depletion of chemicals.
 Following Barrio et al., we chose $\alpha=-\gamma$ in order to have 
 $(0,0)$ as the homogeneous steady state.
 In order to have stable uniform solution, we require either $\alpha \geq 0$
 and $\beta \leq - \alpha$, or $\alpha \leq 0$ and $\beta\leq -1$.
 The ratio of diffusion coefficients between two chemicals is represented
 by $D$, and $\delta$ scales the system size. Cubic
 coupling $r_{1}$ favors striped patterns while quadratic coupling $r_{2}$
 favors spot patterns in flat surface and also on a sphere.
 \paragraph*{}
 The above model was systematically analyzed on two-dimensional flat
 domain~\cite{barrio}. The model was also thoroughly studied on a sphere to understand
 the effect of curvature on patterns~\cite{varea}. Interestingly, these patterns are
  similar to the patterns of silicate formed on the
 membranes of Radiolaria. 
 \section{pattern formation on various non-uniformly curved surfaces}
 In this section, we study pattern formation on various non-uniformly curved
 surfaces. In particular, we analyze the effect of curvature and shape on
 the formation of patterns in these surfaces. Specifically,
 we consider formation of spot and stripe
 patterns on a torus and on an ellipsoid. In order to 
 explore the role of curvature and shape,
 we obtain  patterns on these surfaces by varying the shape parameters (for example, 
 $R$ and $r$ on a torus)
 but keeping the area constant. Our results show that, for the same area,
 changes in the shape can result in different number of spots,
 and 
 can also control the orientation of stripe patterns. 
 \subsection{Torus}
\begin{figure}[h!]
\begin{center}
			
\centering{
\includegraphics[height=3.8cm]{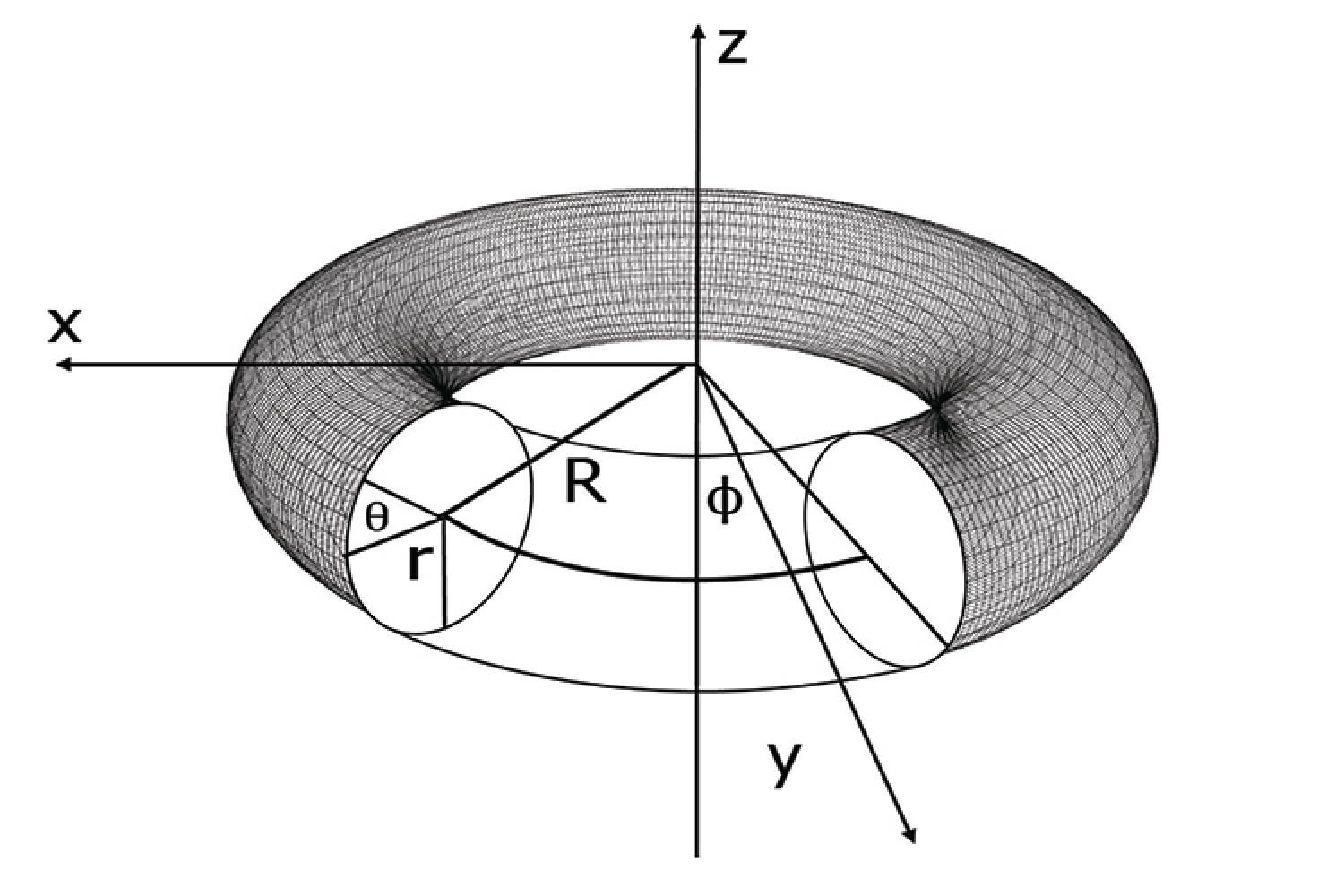}\\  \vspace{-0.5cm}
}
\end{center}
\caption{Schematic representation of the torus parametrized by $(\theta, \phi)$.}
\label{torus1}
\end{figure}

  We now consider the RD equation on a torus and address the following question.
 How does changes in the $R$ and $r$ 
affect the formation of spatial patterns on a torus? Specifically, we study
how the distribution of spots vary as we change the $R$ and $r$
of the torus but keeping area constant. In the case
of stripe patterns, we analyze how the orientation depends on the shape of the torus.
In order to answer above question, we solve numerically RD equation
on a torus for different values of $R$ and $r$ keeping the
area same in all cases, and thus explore the role of curvature and shape on the 
formation of spatial patterns.
\paragraph*{}
 The surface of a torus can be parametrized as
 \begin{equation}
	X(\theta,\phi) = 
		\begin{pmatrix} 
			(R+r\cos \theta)\cos \phi \\ 
			(R+ r\cos \theta)\sin \phi \\ 
			r\sin \theta
		\end{pmatrix},
		\label{eq:3}
\end{equation}
and the coordinates $\theta$ and $\phi$ as well as the major radius $R$
and minor radius $r$ can be visualized as in Fig.~\ref{torus1}. Here both 
coordinates $\theta$ and $\phi$ vary from $0$ to $2\pi$.
The Gauss curvature $K$ and mean curvature $H$
of the torus can be read as (see appendix)
\begin{eqnarray}
\begin{aligned}
 K = \frac{\cos\theta}{r(R+r\cos\theta)},\\
 H=\frac{-(2r\cos\theta+R)}{2r(R+r\cos\theta)},
 \end{aligned}
 \label{eq:4}
\end{eqnarray}
and note that on a torus both 
these curvatures are $\theta$-dependent. Gauss curvature is positive
on the outer side of the torus, and negative on the inner side. Note that
the Gauss curvature vanishes at $\theta=\pi/2$ and $\theta=3\pi/2$. The mean
curvature is negative from $0$ to $\pi/2$, and depending on $R$ and $r$,
it can be positive or negative from $\pi/2$ to $\pi$.

\paragraph*{}
 In order to solve Eq.~(\ref{eq:rd_barrio}) numerically on the surface of 
 a torus, we consider the Laplace-Beltrami 
  operator $\bigtriangleup_{LB}$ on the surface of a torus. The 
  Laplace- Beltrami operator is given in curvilinear 
  coordinates as 
  \begin{equation}
   \bigtriangleup_{LB}=\sum_{i,j=1}^{2}\frac{\partial}{\partial q^{i}}(\sqrt{g}g^{ij}
   \frac{\partial}{\partial q^{j}}),
   \label{eq:5}
  \end{equation}
 with $g=\det(g_{ij})$, where $g_{ij}$ is the metric elements on the surface.
 The Laplace-Beltrami
operator on a torus is explicitly given by
\begin{equation}
 \bigtriangleup_{LB}=\frac{1}{(R+r \cos\theta)^{2}}\frac{\partial^{2}}{\partial \phi^{2}}
 +\frac{1}{r^{2}}\frac{\partial^{2}}{\partial \theta^{2}}-\frac{\sin\theta}{r(R+r~\cos\theta)}
 \frac{\partial}{\partial \theta},
 \label{eq:6}
\end{equation}
 and the Eq.~(\ref{eq:rd_barrio}) is then numerically solved using the explicit Euler method, where coordinates $\theta$
 and $\phi$ are discretized as $\theta_{m}=m\bigtriangleup\theta$, where $m=0,1,\ldots,(M-1)$, and 
 $\phi_{n}=n\bigtriangleup\phi$, where $n=0,1,\ldots,(N-1)$. We use periodic
 boundary conditions for both $\theta$ and $\phi$. We set the initial condition choosing
 the random values between -0.5 and 0.5 on a circle near $\phi=\pi/2$, and all other
 points we take $u=v=0$.
\begin{figure}[h!]
\begin{center}
	
\centering{
\includegraphics[height=4.5 cm]{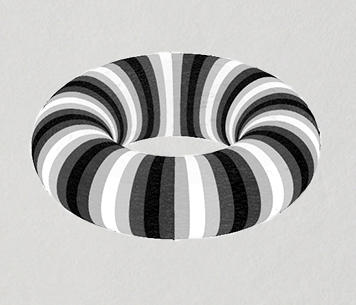}\\  \vspace{-.5cm}
}
\end{center}
\caption{Pattern of u on a torus with parameters  $D=0.516$, $\alpha=0.899
$, $\beta=-0.91$, $\gamma=-\alpha$, $\delta=0.0045$, $r_{1}=3.5$, $r_{2}=0$ with $R=1$, $r=0.1$.}
\label{torus2}
\end{figure}
 \paragraph*{}
 To begin with, the pattern on a torus with parameters specified in the caption
 of Fig.~\ref{torus2} is obtained with $2\pi r<\lambda_{T}$,
 where $\lambda_{T}$ is the Turing length. 
 Since $2\pi r<\lambda_{T}$, the above case is similar to pattern formation
 on a cylinder with radius $r$ and height $2 \pi R$ . As expected we obtain $10$ rings on the torus.
 In this case taking different $R$ and $r$ values with $2\pi r<\lambda_{T}$ only 
 changes the number of rings on the torus. Here the ring-like 
  shape of the pattern will not change as we vary $R$ and $r$. Thus the shape 
  of the torus will not produce any qualitative changes on patterns when $2\pi r<\lambda_{T}$ . 
  In order to see the effect of curvature and domain shape
 on patterns, we consider $2\pi r >\lambda_{T}$, and then 
  produce different patterns on a torus by changing $R$ and $r$ values.
  \begin{figure}[h!]
  \centering
  \begin{minipage}{.2 \textwidth}
   \centering
   \includegraphics[height=40mm,width=40mm]{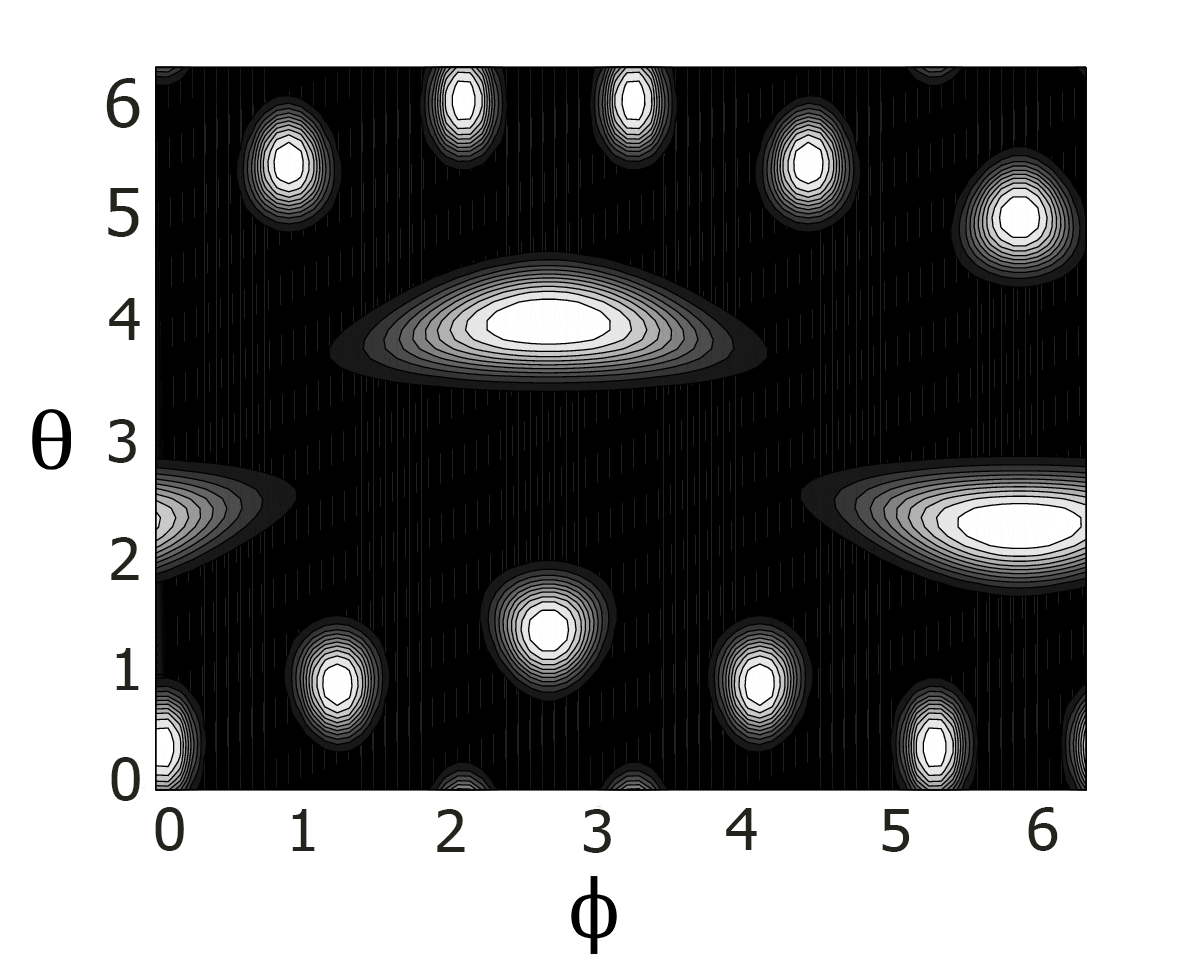}
  \end{minipage}
  \hspace{1em}
\begin{minipage}{.2 \textwidth}
 \centering
 \includegraphics[height=40mm,width=40mm]{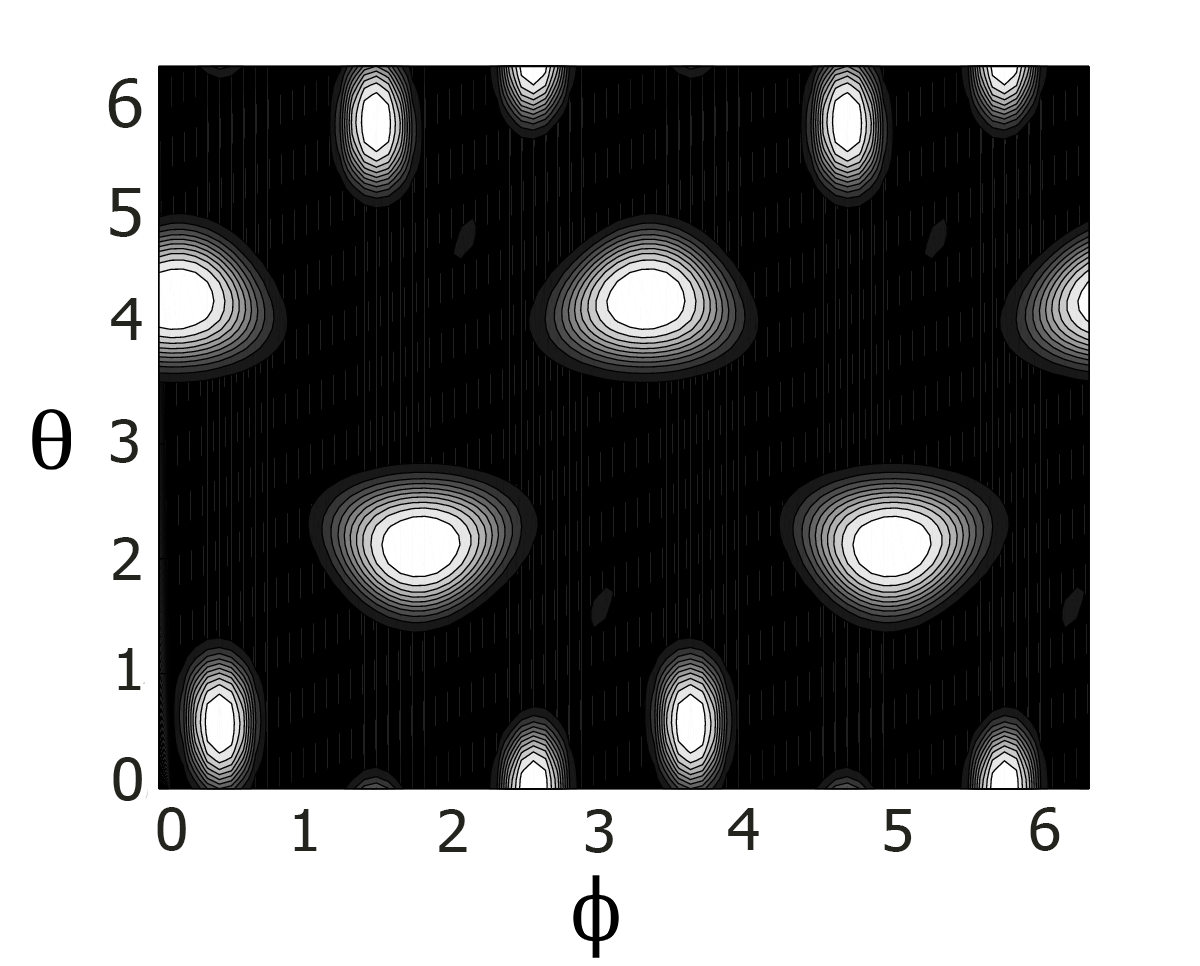}
\end{minipage}
\caption{ Left is the 12 spots on a torus with 
$R=0.6$, $r=1/0.6\pi$ and right is the 10 spots with 
 $R=0.7$, $r=1/0.7\pi$. The parameter values are $D=0.516$, $\alpha=0.899
$, $\beta=-0.91$, $\gamma=-\alpha$, $\delta=0.0085$, $r_{1}=0.02$, $r_{2}=0.2$. Vertical
axis is $\theta$ and $\phi$ is horizontal axis. }
\label{spots_on_torus}
 \end{figure}
 
\begin{figure}[h!]
\begin{center}
	
\centering{
\includegraphics[height=4.5 cm]{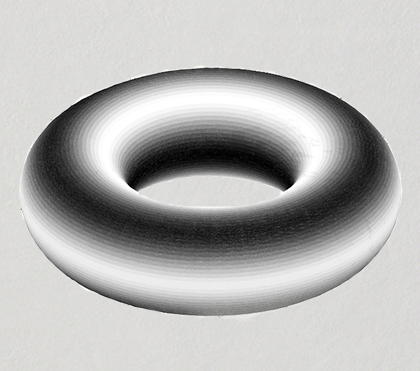}\\  \vspace{-.5cm}
}
\end{center}
\caption{Stripe pattern on a torus with parameters  $D=0.516$, $\alpha=0.899$,
$\beta=-0.91$, $\gamma=-\alpha$, $\delta=0.0045$, $r_{1}=3.5$, $r_{2}=0$ with
$R=0.90946$, $r=0.35$.}
\label{torus_ring}
 \end{figure}

  \paragraph*{}
  First we consider formation of spot patterns on the torus. We choose the parameter values
$r_{1}=0.02$, $r_{2}=0.2$, which result in spot patterns on flat surface,
and on sphere. We obtain patterns with 12 spots and 10 spots on a torus,
as shown in Fig.~\ref{spots_on_torus}, for two cases respectively both with same 
area and initial
condition but with different values of shape parameter. This result thus shows that the 
number of spots can  be influenced by the shape. Here, the
torus with smaller $R$ has more number of spots compared to the torus with 
 large $R$. Note that in both cases number of spots 
 are more in the region $0<\theta<\pi/2$ compared to the region $\pi/2<\theta<\pi$.
 \paragraph*{}
  We also observe that the number of spots in region
 $0<\theta<\pi/2$ is more in the case of a torus with smaller~ $R$. Specifically,
  torus with smaller $R$ has 7 spots 
  in this region but 
  the torus with higher $R$ contain only 6 spots. But in the region between $\pi/2$
  and $\pi$ the torus with smaller $R$ has one spot, and
  the torus with larger $R$ contain two spots. The size of the spots are larger in
the region between $\pi/2$ and $3\pi/2$ in both cases of higher
and smaller R. Note that the smaller R has larger spots
in this region compared to higher R.
  \paragraph*{}
  In both cases, outer side of the torus contain more number of spots than inner side.
  These results thus illustrate that specific values of shape parameters  
 can result in localization of chemicals, where the concentration can be more
 on outer side 
 of the torus than inner side. Note that the quadratic term $r_{2}$
favors spot patterns on the surface of a torus like in flat and spherical geometry. 
  \begin{figure}[h!]
\begin{center}
	
\centering{
\includegraphics[height=4.5 cm]{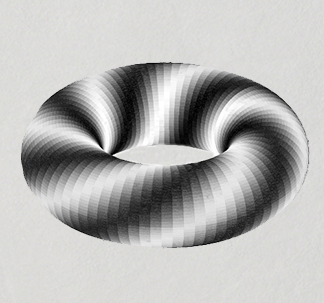}\\  \vspace{-.5cm}
}
\end{center}
\caption{Stripe pattern on a torus with parameters  $D=0.516$, $\alpha=0.899$,
$\beta=-0.91$, $\gamma=-\alpha$, $\delta=0.0045$, $r_{1}=3.5$, $r_{2}=0$ with
 $R=1$, $r=0.3$ .}
 \label{torus_stripe}
 \end{figure}
\paragraph*{}
Next we study stripe patterns on a torus. We obtain different orientation of stripe
patterns by changing  $R$ and $r$ as shown in Fig.~\ref{torus_stripe} and Fig.~\ref{torus_ring}. 
In the first case,
we produce stripe pattern wrapping around the torus. In the second case,
with same initial condition and parameters as in the first case, we obtain ring-like pattern
where the concentration is varying only along $\theta$- direction by changing 
$R$ and $r$.
Thus changes in the shape can result in different
 orientation of stripe patterns on toroidal surfaces. Note that the cubic term $r_{2}$
favors stripe  patterns on the surface of a torus like in the flat and the spherical geometry.
\paragraph*{}
 An intuitive understanding of our results can be given in the following way. As described 
 in~\cite{nucleation}, the Laplace-Beltrami operator $\bigtriangleup_{LB}$ on 
 a torus can be mapped to as Laplace operator on a flat surface  with a conformal
 factor, and is given as
 \begin{equation}
  \bigtriangleup_{LB}=\frac{(\cosh\eta-\cos \theta_{i})^{2}}{a^{2}}
  \left(\frac{\partial^{2}}{\partial \theta_{i}^{2}}+\frac{\partial^{2}}{\partial \tilde{\phi}^{2}}
  \right),
  \label{eq:7}
 \end{equation}
where 
\begin{eqnarray*}
a &=&(R^{2}-r^{2})^{1/2},\\
\eta&=&\coth^{-1}\left(\frac{R}{\sqrt{R^{2}-r^{2}}}\right),\\
\tilde{\phi}&=&
\phi \sinh\eta, \\
\cos\theta_{i}&=&
\cosh\left[\coth^{-1}\left(\frac{R}{\sqrt{R^{2}-r^{2}}}\right)\right]-\frac{a \sinh\frac{R}{\sqrt{R^{2}-r^{2}}}}{r\cos\theta
+R}. 
\label{eq:8}
\end{eqnarray*}
Thus the RD equation on a torus can be equivalently described
as a RD equation on a flat surface where $\delta$ is replaced by $\theta$-dependent
$\delta_{eff}$,
where $\delta_{eff}=\delta
(\cosh\eta-\cos \theta_{i})^{2}/a^{2}$. Hence, the effect of curvature is
equivalent to having a spatially varying parameter $\delta_{eff}$.
Since the variations 
in the $\delta_{eff}$ depends on $R$ and $r$, the orientation of stripe pattern as well as 
the size and number of spots can vary
as we change the shape of the torus. 
\begin{figure}[h!]
\begin{center}
			
\centering{
\includegraphics[width=5cm]{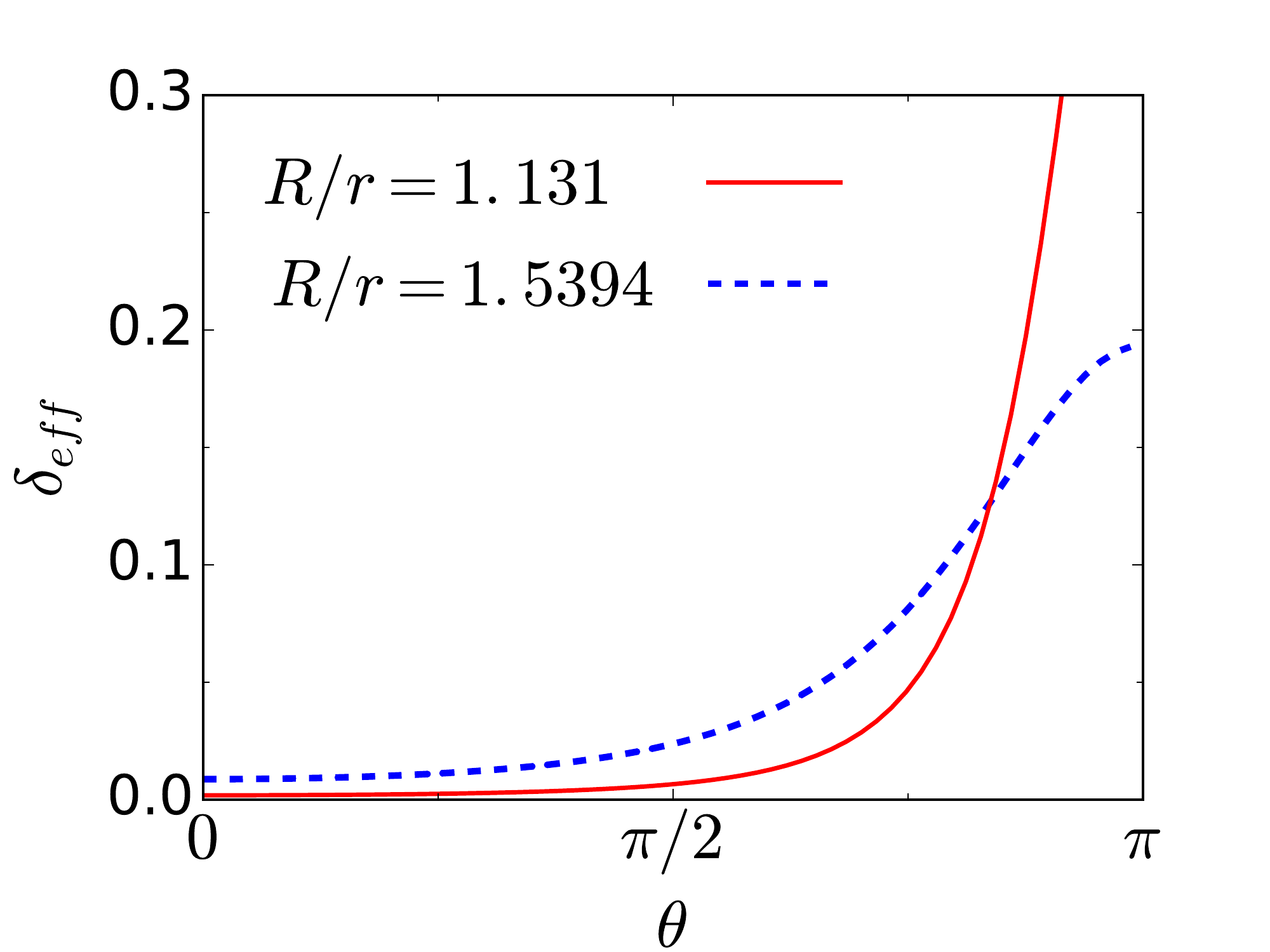}\\  
}
\end{center}
\caption{Figure illustrates the variation of $\delta_{eff}$ as the coordinate
$\theta$ changes.}
\label{effective_delta}
\end{figure}
\paragraph*{}
Here we illustrate 
the effect of local variation in $\delta$ 
by considering the example of RD on a cylinder.
 We consider the RD equation on a cylinder with parameters as specified
 in the caption Fig.~\ref{torus_stripe}, with 
radius~$r=0.3$ and height $h=2\pi$. In this case, we have obtained
an inhomogeneous concentration along $\theta$-direction on a cylinder, 
 but stripe pattern will not wrap around the cylinder. The stripe pattern on the 
 cylinder thus clearly differs from the 
 the stripe pattern that wraps on the torus with same initial condition. Now 
suppose we assume two different values
in the parameter $\delta$ on a cylinder, where $\delta=0.0085$ for
$0\leq \theta \leq (M/2-1)2\pi/M$ and $\delta=0.0045$ 
for $\pi\leq\theta\leq(M-1)2\pi/M$, and with same initial condition as above. In this case 
we obtain a stripe pattern with varying concentration along the 
of $\theta$-direction, where the stripe
wraps around the cylinder. This 
example thus illustrate that the wrapping of stripe pattern on a torus
is due to curvature, and whose effect is 
equivalent to having a spatial-dependent $\delta_{eff}$, 
and not due to a specific initial condition.
\paragraph*{} 
Suppose we now change $\delta$, say to $0.0085$ for 
$0\leq \theta \leq (M/2-1)2\pi/M$ and to $0.011559$ for $\pi\leq\theta\leq(M-1)2\pi/M$.
Then we obtain the pattern on a cylinder
which is qualitatively same as the pattern obtained on the second torus in 
the Fig.~\ref{torus_ring}, where 
it forms a ring-like pattern with variation along $\theta$- direction. Thus, different
local variations in $\delta$ can give different orientations to the stripe
pattern. Thus the changes in the shape can give different spatial
variations in $\delta$, which in turn can lead to 
different orientation of stripe patterns.
\paragraph*{}
The localization of chemicals in preferred regions and an increase
in the size of the spots in certain regions can also be intuitively
understood in the following way. Note from the Fig.~\ref{effective_delta}, the $\delta_{eff}$
increases from $\theta=0$ to $ \pi$, but the gradient in $\delta_{eff}$ is large
in the region $\pi/2$ to $\pi$. Since the Turing wavelength increases with increase in $\delta$,
in this picture 
one can intuitively argue that 
 an increase in $\delta_{eff}$ can lead to higher wavelength mode in the
region $\pi/2$ to $\pi$ compared to $\theta=0$ to $ \pi/2$.
Since the gradient in $\delta_{eff}$ is large and positive
in the region $\pi/2$ to $ \pi$ on a torus with both smaller and larger $R$, 
 one can expect 
less number of spots with larger size in the region $\pi/2$ to $ \pi$ compared
to that between $0$ to $ \pi/2$ in both cases.  Note that 
the large gradient in $\delta_{eff}$ can result in the localization of chemicals
 and larger size spots in some preferred regions.
\paragraph*{}
To sum up, varying $R$ and $r$ 
 on the torus can lead to different $\delta_{eff}$ which in turn can control the  
 size and distribution of spots, and can also lead to different 
 orientation of stripes. Our results also indicate that by controlling the curvature
 and shape one can 
 drive the chemicals to a preferred region. This result may useful in the studies 
 of self-organization of molecules in biological membranes.
 \subsection{Ellipsoid}
\begin{figure}[h!]
  \centering
  \begin{minipage}{.2 \textwidth}
   \centering
   \includegraphics[height=40mm,width=40mm]{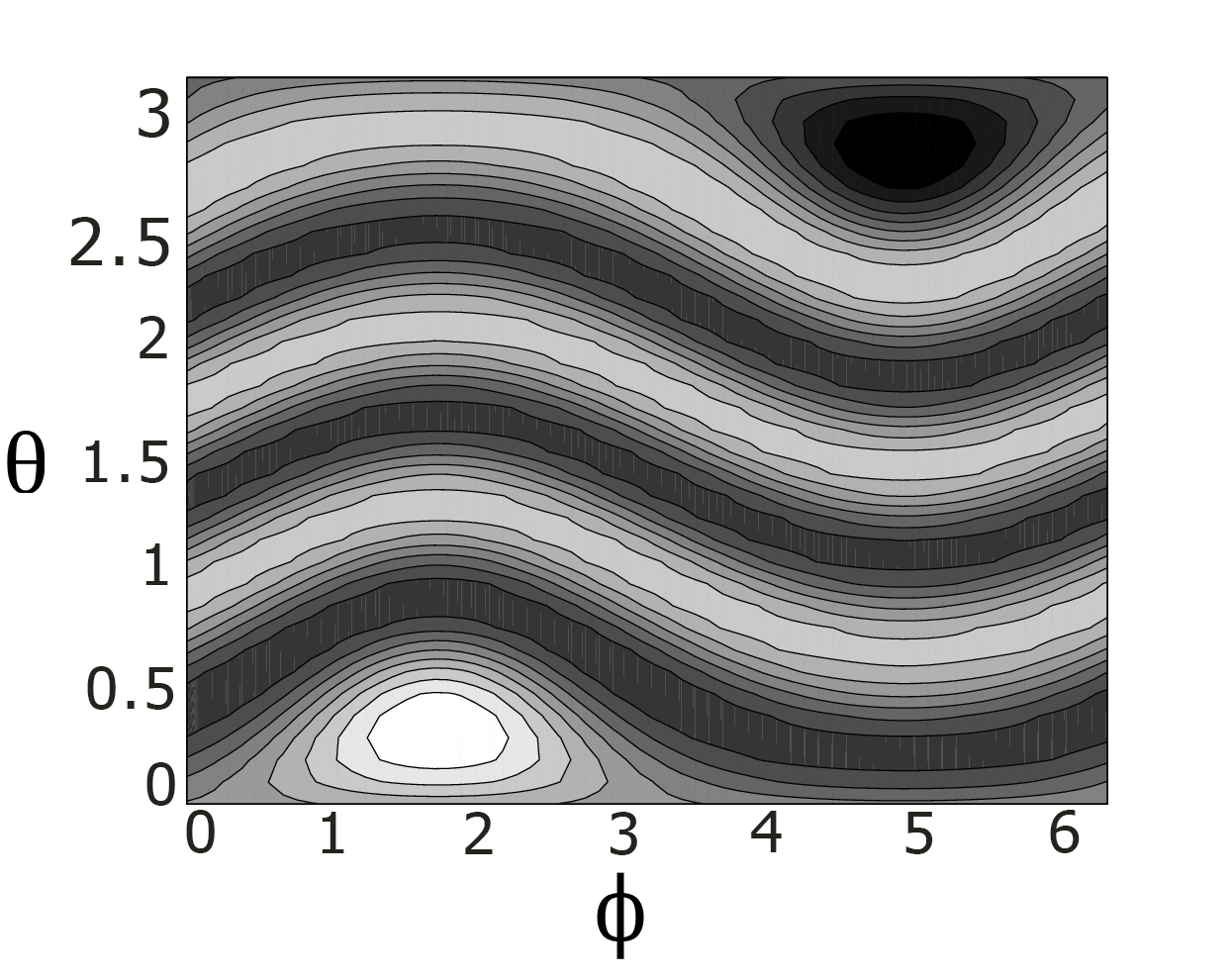}
  \end{minipage}
  \hspace{1em}
\begin{minipage}{.2 \textwidth}
 \centering
 \includegraphics[height=40mm,width=40mm]{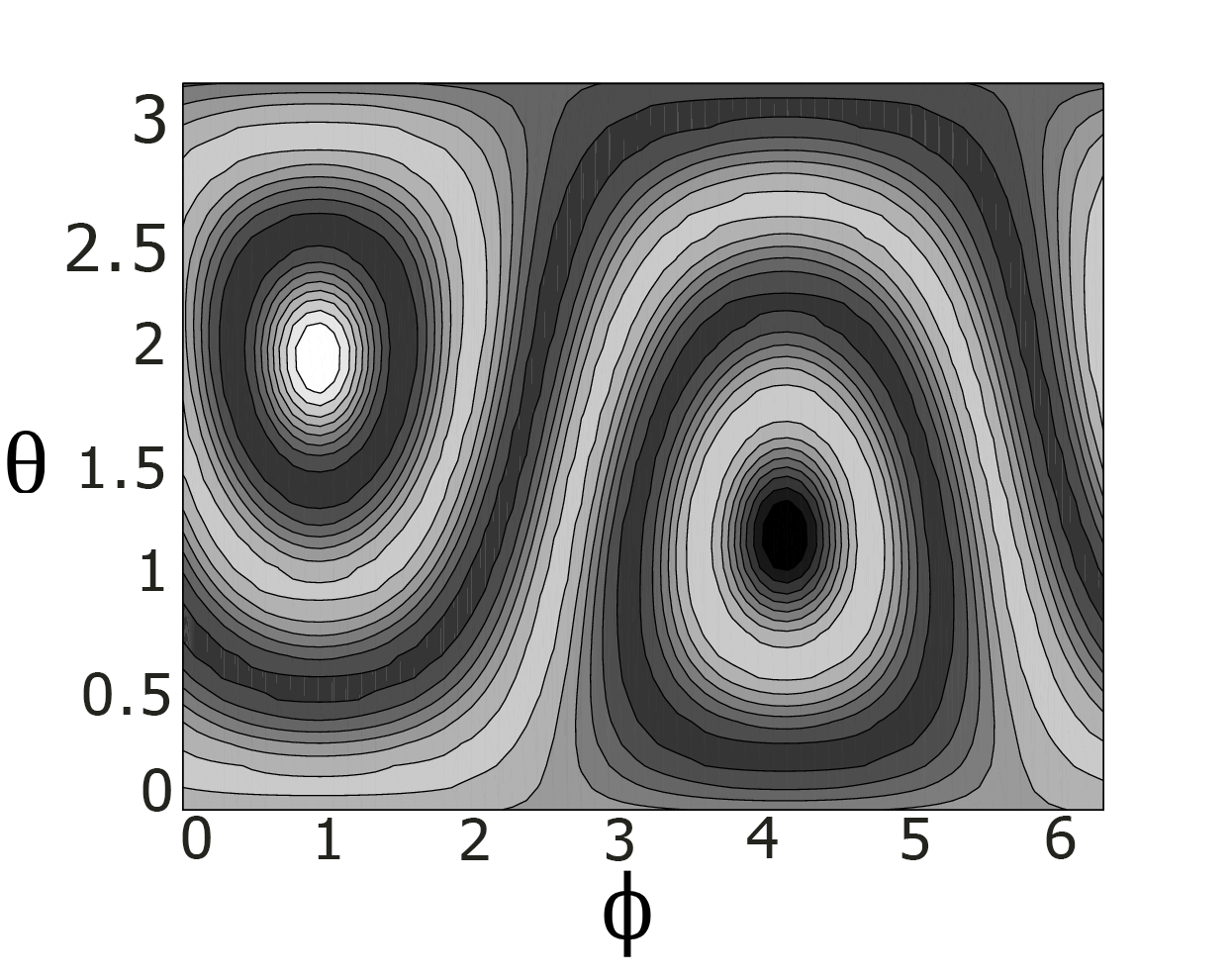}
\end{minipage}
\caption{ Left is the stripe pattern on a prolate spheroid with $b=1.2, a=1$ and 
right is on an oblate spheroid $b=1$, $a=1.09884$. The parameter values are $D=0.516$, $\alpha=0.899
$, $\beta$=-0.91, $\gamma=-\alpha$, $\delta~=0.0085$, $r_{1}=0.02$, $r_{2}=0$.}
\label{stripe_ellipse}
 \end{figure}

  We now consider the pattern formation on an ellipsoid.
Specifically, we obtain patterns on oblate and prolate
spheroid with same area, and analyze the role of 
curvature and shape parameters on the number of spots,
and on the orientation of stripe pattern. 
 The equation of an ellipsoid is
  \begin{equation}
   \frac{x^{2}+y^{2}}{a^{2}}+\frac{z^{2}}{b^{2}}=1,
   \label{eq:9}
  \end{equation}
where the case with $a>b$ is 
called oblate spheroid, while the case with $a<b$ is prolate spheroid.
The ellipsoid can be parametrized as
 \begin{equation}
	X(\theta,\phi) = 
		\begin{pmatrix} 
			a\sin\theta\cos\phi \\ 
			a\sin\theta \sin\phi \\ 
			b \cos\theta
		\end{pmatrix},
		\label{eq:10}
\end{equation}
where $\theta$ and $\phi$ are the coordinates on the surface. 
Note that on an ellipsoid both curvatures are $\theta$- dependent. The Gaussian curvature
of an ellipsoid is positive (see appendix) where the 
curvature varies from $b^{2}/a^{4}$ (at $\theta=0$) to 
$1/b^{2}$ (at $\theta=\pi/2$).
 \paragraph*{}
 We solve Eq.~(\ref{eq:rd_barrio}) numerically on the surface of an ellipsoid.
The Laplace-Beltrami 
operator $\bigtriangleup_{LB}$ on an ellipsoid can be read from $g_{ij}$ (see appendix) and given as
\begin{eqnarray*}
\bigtriangleup_{LB}=\frac{1}{(a^{2}\cos^{2}\theta+b^{2}\sin^{2}\theta)}\frac{\partial^{2}}{\partial \theta^{2}}
 +\frac{1}{a^{2}\sin^{2}\theta}\frac{\partial^{2}}{\partial \phi^{2}}\nonumber\\+
 (\frac{\cot\theta}{(a^{2}\cos^{2}\theta+b^{2}\sin^{2}\theta)}+
 \frac{(1-b^{2}/a^{2})\sin2\theta}{2 a^{2}(\cos^{2}\theta+b^{2}/a^{2}\sin^{2}\theta)^{2}})
 \frac{\partial}{\partial \theta},
 \label{eq:11}
 \end{eqnarray*}
and the co-ordinates on an ellipsoid can be discretized as
 $\theta_{m}=(m+1/2)\bigtriangleup\theta$, where $m=0,1,\ldots,(M-1)$, and 
 $\phi_{n}=n\bigtriangleup\phi$, where $n=0,1,\ldots,(N-1)$. Following~\cite{varea},
 we consider the nearest neighbor of u$(m=0,M;n=0,1,\ldots,N/2-1)$ is u$(m=0,M;N/2,\ldots,N-1)$, 
 and for $\phi$ co-ordinate we use periodic boundary condition. The same condition
 is taken for v also. We choose initial condition as
 random values between -0.5 and 0.5 on a circle near equator, and all other
 points we take $u=v=0$.
\begin{figure}[h!]
  \centering
  \begin{minipage}{.2 \textwidth}
   \centering
   \includegraphics[height=40mm,width=40mm]{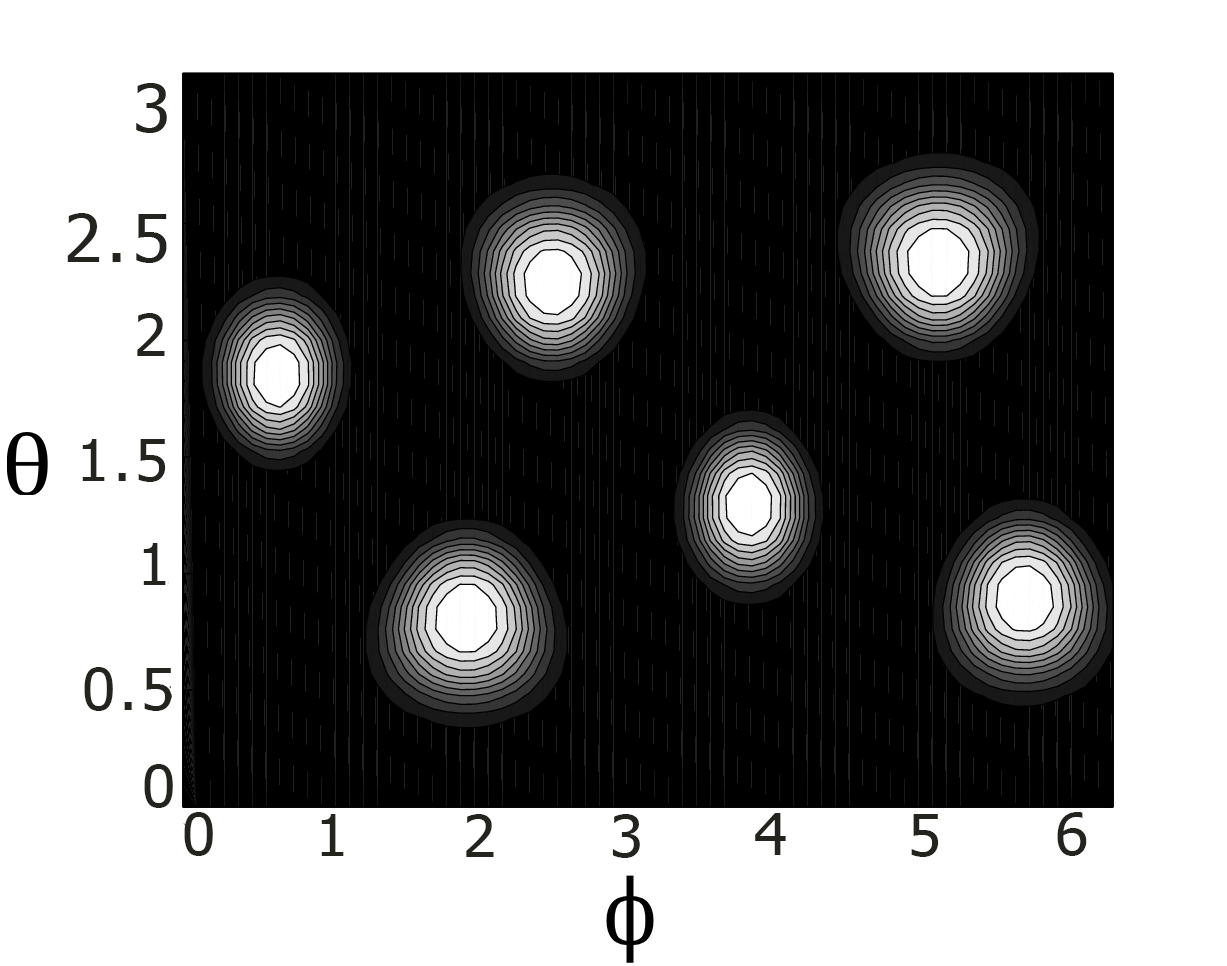}
  \end{minipage}
  \hspace{1em}
\begin{minipage}{.2 \textwidth}
 \centering
 \includegraphics[height=40mm,width=40mm]{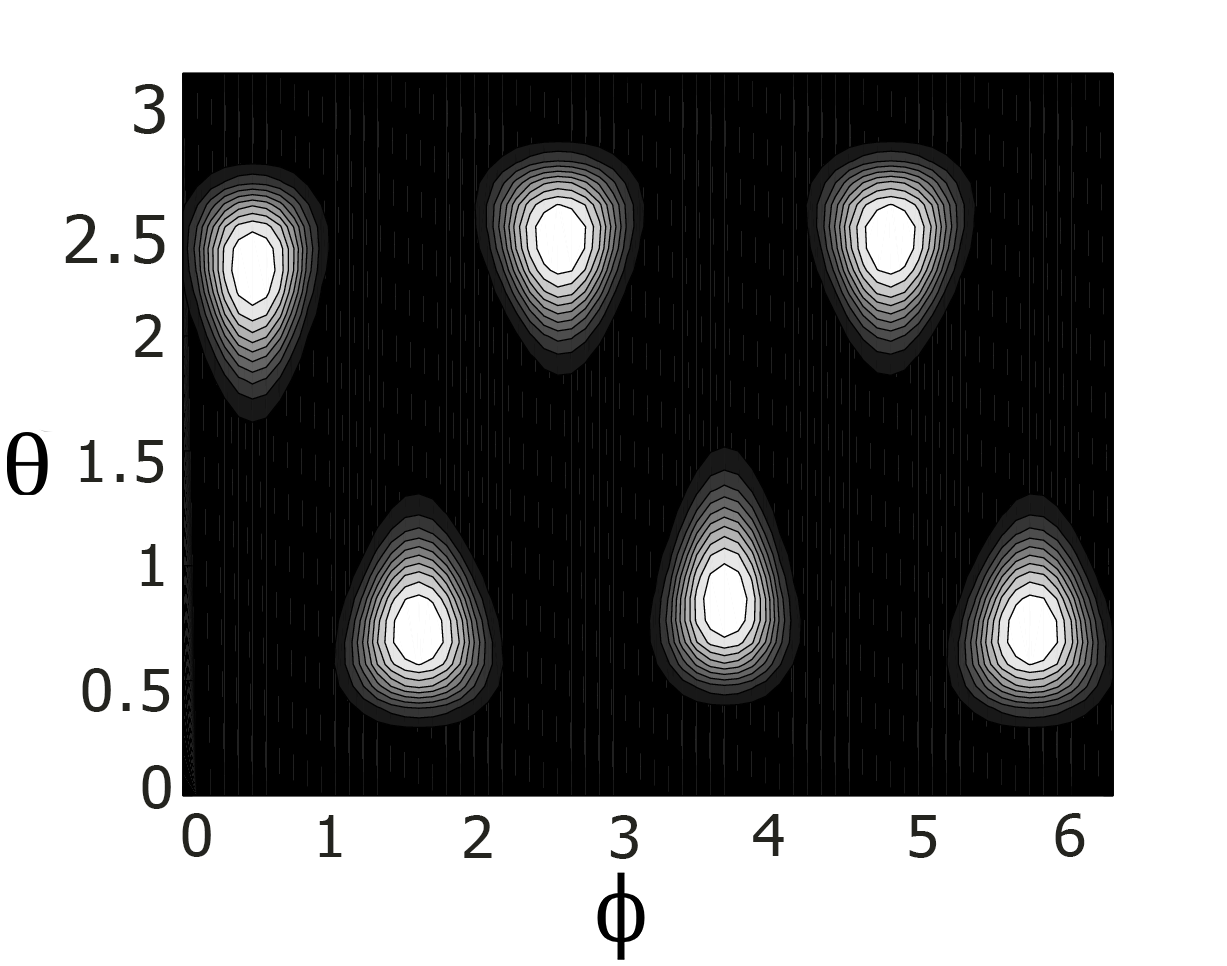}
\end{minipage}
\caption{ Left is the 6 spots on a prolate spheroid with 
$b=1.1$, $a=0.9$ and right is the 6 spots on an oblate spheroid
of same area with $b=0.50493$, $a=1.2$. The parameter values are $D=0.516$, $\alpha=0.899
$, $\beta=-0.91$, $\gamma=-\alpha$, $\delta=0.0171$, $r_{1}=0.02$, $r_{2}=0.2$.}
\label{spot_ellipsoid_one}
 \end{figure}
\paragraph*{}
First we consider formation of spot patterns. We obtain  
6 spots on an oblate spheroid and 6 spots on prolate spheroid with same initial condition
and area but with different values of $b$ and $a$ as shown in Fig.~\ref{spot_ellipsoid_one}.
Note that 
the position of spots are different in both cases. Thus it is clear that the elongation
of an ellipsoid can control the position of spot patterns.
Furthermore, numerical simulations give 6 spots when $b=0.99$, $a=1$ as 
shown in Fig.~\ref{spot_sphere}. This is exactly same as  
the previous result obtained on a sphere~\cite{varea} with same parameters.
The above result illustrates that small deviations from uniform
curvature cannot
influence pattern formation as pointed out in ~\cite{venkataraman}.
\paragraph*{}
 It is interesting to note that position and number of spots
on a prolate spheroid with $b=1.1$ and $a=0.9$ is exactly same as that 
on a sphere with $b=1$ and $a=1$. But the position of spots on the oblate spheroid 
is different from that on a sphere. Here, the effect of curvature and 
shape is more pronounced in the case of an oblate spheroid compared to the prolate one.
Note that the quadratic term $r_{2}$
favors spot patterns on the surface of an ellipsoid like in the other geometries considered.
\begin{figure}[h!]
\centering
\includegraphics[height=40mm,width=55mm]{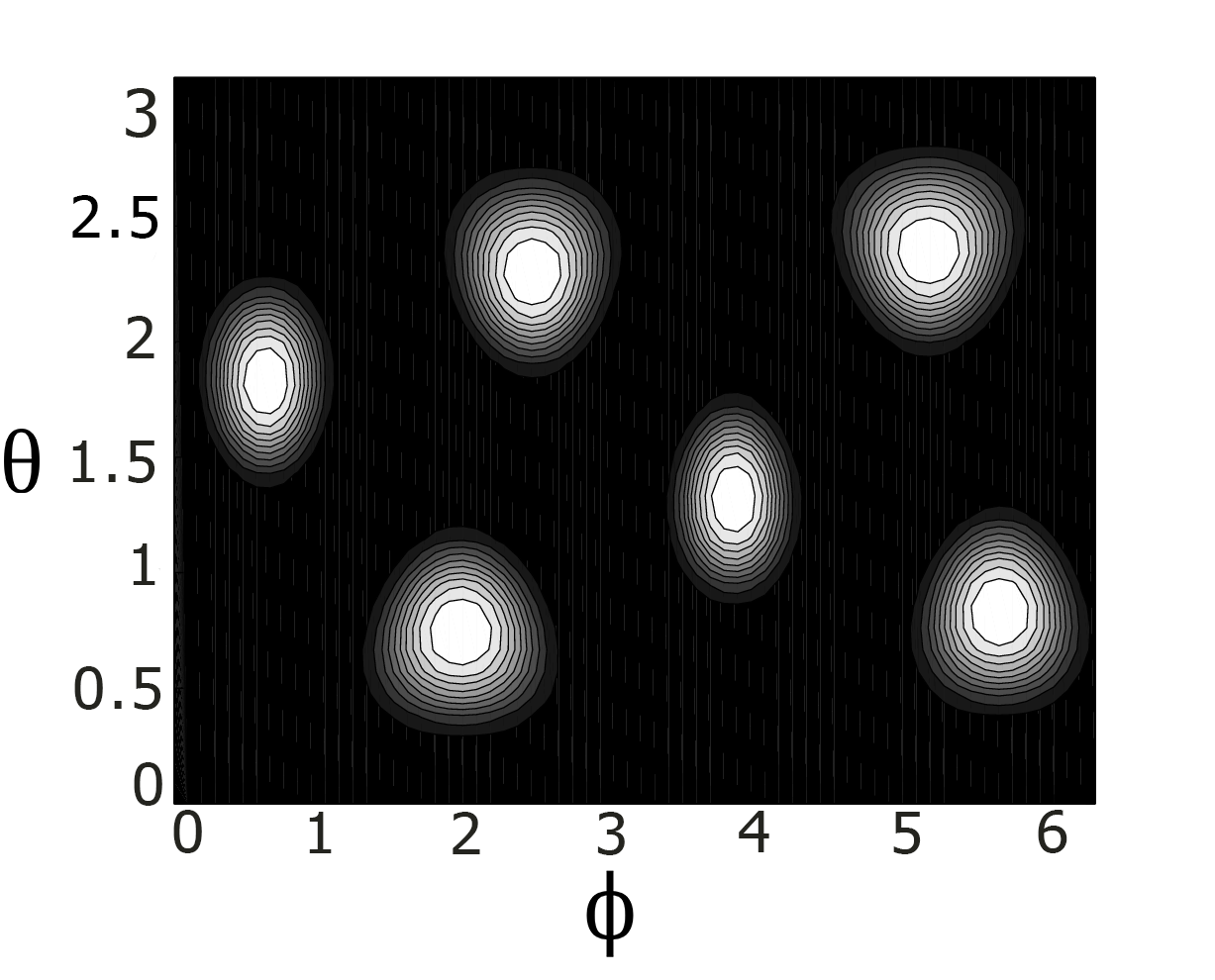}
\caption{6 spot pattern with $b=0.99$, $a=1$. Other parameters are same as in 
Fig.~\ref{spot_ellipsoid_one}.
}
\label{spot_sphere}
\end{figure}
\begin{figure}[H]
  \centering
  \begin{minipage}{.2 \textwidth}
   \centering
   \includegraphics[height=40mm,width=40mm]{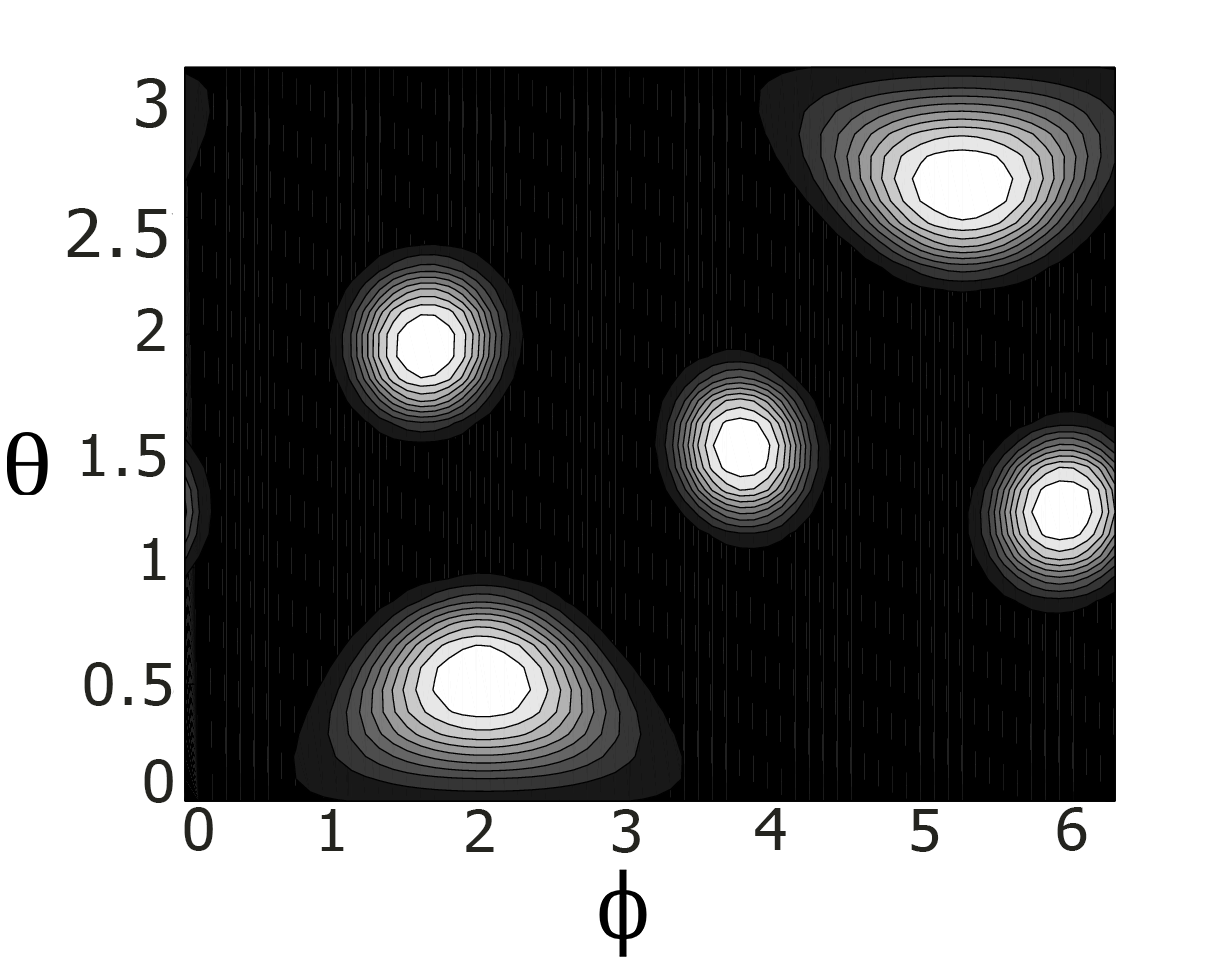}
  \end{minipage}
  \hspace{1em}
\begin{minipage}{.2 \textwidth}
 \centering
 \includegraphics[height=40mm,width=40mm]{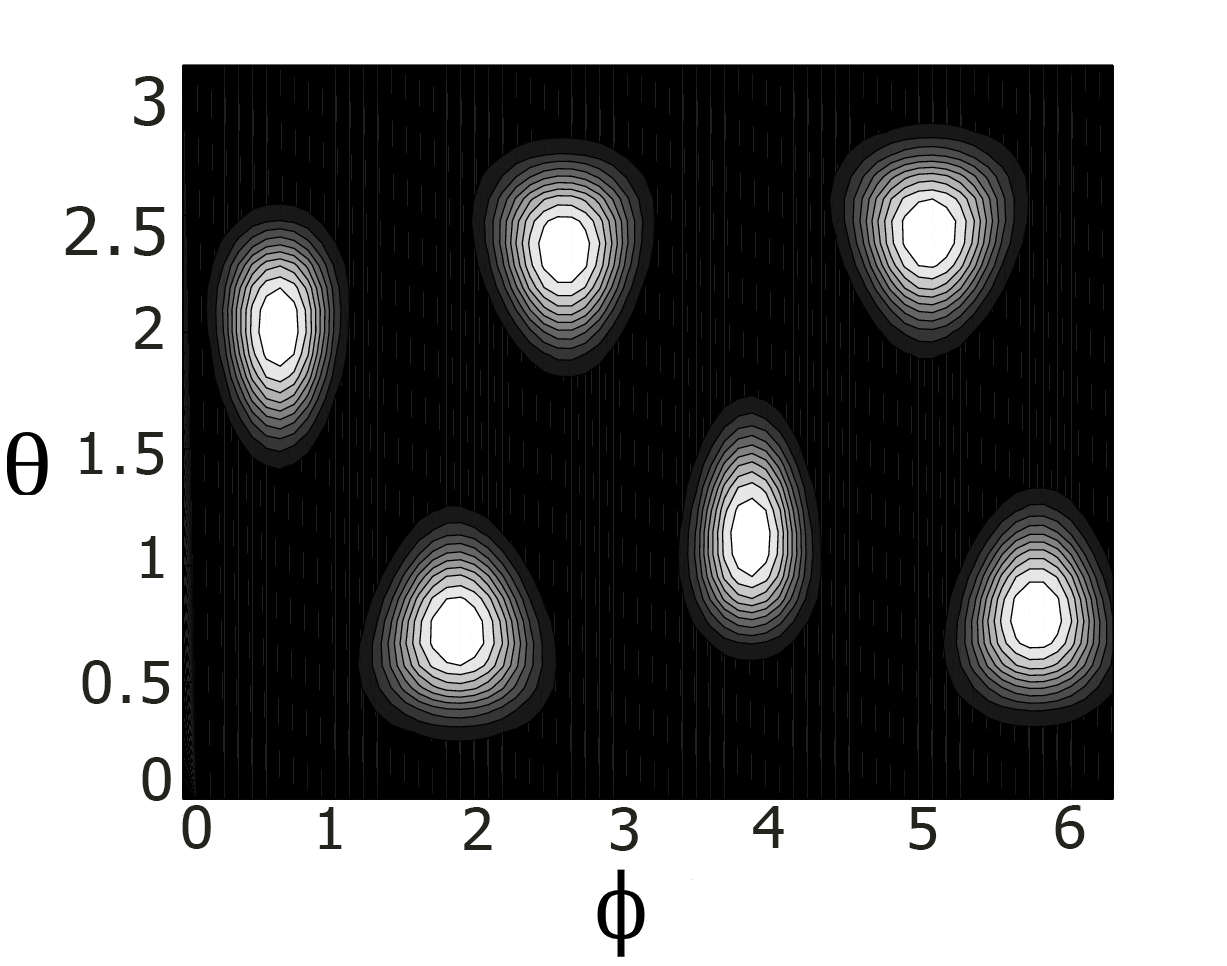}
\end{minipage}
\caption{ Left is the 5 spots on a prolate spheroid with 
$b=1.10409$, $a=0.8$ and right is the 6 spots on an oblate spheroid
of same area with $b=0.7$, $a=1$. The parameter values are $D=0.516$, $\alpha=0.899
$, $\beta=-0.91$, $\gamma=-\alpha$, $\delta=0.0171$, $r_{1}=0.02$, $r_{2}=0.2$.}
\label{spot_ellipsoid_two}
 \end{figure}
 \paragraph*{}
 As shown in the Fig.~\ref{spot_ellipsoid_two}, we again produce spots where the parameters
 considered are same as in the caption of Fig.~\ref{spot_ellipsoid_one}, but with different shape.
 We have then obtained different number of spots on a prolate and an oblate spheroid.
 Specifically oblate spheroid has more number of spots
 compared to the prolate one. In this case, note that
  the size of spots below $\theta=\pi/4$ and above $\theta=3\pi/4$ are larger on the prolate 
 spheroid. Thus the effects of the curvature and domain shapes are more 
 pronounced in this case compared to the previous one given in the Fig.~\ref{spot_ellipsoid_one}.
\paragraph*{}
Next we consider the formation of stripe patterns. Fig.~\ref{stripe_ellipse} shows stripe patterns
on a prolate spheroid and an oblate spheroid of same surface
area. Note that the orientation of
stripe patterns are different in both cases. Hence 
 the elongation of an ellipsoid can influence the orientation of stripe patterns.
Similar to other geometries considered earlier, even 
in the case of ellipsoid the cubic term $r_{1}$
favors stripe patterns.
    \section{conclusion}
\paragraph*{}
To sum up, we have studied pattern formation on various non-uniformly curved surfaces.
In particular, we analyzed the role of curvature and shape on the pattern formation.
First we obtained
spot and stripe patterns on a torus. To begin with, we have produced patterns on a torus 
similar to that on a cylinder by considering $2\pi r< \lambda_{T}$. We have then relaxed
the above condition, and obtained spot and stripe patterns for different
$R$ and $r$ values. For the same area, our results 
indicate that the orientation of stripe pattern as well as 
the size and number of spots can be controlled by the shape of the torus.
\paragraph*{}
Our analysis on a torus shows that the number of spots can be controlled by 
the major and minor radii $R$ and $r$ of the torus, respectively. 
Secondly the curvature and domain shape can also control 
the size of the spots. For the parameter values considered, the number of spots
on the outer side of the torus in each case is found to be more than that on the inner side,
the difference in number being dependent on the gradient in the parameter $\delta_{eff}$.
We also show that it is possible to produce different orientations of stripe patterns
by changing $R$ and $r$. Thus our results indicate that curvature
and shape can play an important role in the formation of patterns on non-flat geometries.
\paragraph*{}
We have also suggested an intuitive way of understanding the effect of curvature and shape. 
On a torus, the effect of curvature is equivalent to replacing the parameter
$\delta$ by a space-dependent parameter $\delta_{eff}$. The parameter
$\delta_{eff}$ can vary only along $\theta$- direction and depends
on the shape parameters. Thus the shape can control the formation of patterns.
Note that from the $\delta_{eff}$ one can identify outer side of the torus
can have more number of spots compared to inner side. 
The conformal factor also 
can be useful to understand the size of the spots 
in different regions. Hence calculating confor-
mal factor either analytically or numerically on various
curved surfaces can be important in RD studies.
\paragraph*{}
 We then studied spot and stripe patterns on both prolate and oblate spheroid.
 We show that the number and 
 position of spot patterns can depend on the elongation of an ellipsoid. 
 In one case, we have obtained less number of spots on a prolate spheroid, while 
 in other case both prolate and oblate spheroid contains same number of spots.
 Our analysis shows that shape parameters can influence both position
 as well as the size and number of spots on an ellipsoid similar to that on a torus.
 We also showed that the difference in shape (oblate or prolate spheroid) can 
 result in different orientation of
 stripe patterns. 
 \paragraph*{}
 It is clear from our analysis that the curvature effect is equivalent to
 having an effective anisotropy
 in local parameters
 that can lead to new effects on patterns that it is different from flat case.
 The effect of anisotropy in local parameters is studied in some 
 of the previous woks~\cite{meron1,benson}.
 For instance, effect of anisotropy in diffusion coefficient 
 on flat surface is analyzed~\cite{benson}, and showed that
 the resulting pattern can have
 spatially varying amplitude and wavelength. 
 This is similar to our observations, for example, on a torus where the distance
 between the spots and their size vary with position.
 Similarly, features like stratification 
 of labyrinthine patterns due to anisotropic diffusion mentioned in the work~\cite{meron1}
  can also arise in RD systems on curved surfaces.
 \paragraph*{}
  Some of the earlier studies
 explained the directionality of stripe patterns on fishes by introducing
 anisotropic diffusion coefficients in RD equations on a flat surface~\cite{shoji,shoji1}. 
 Since the curvature can control the orientation of stripe pattern,  
 it would be imperative to incorporate the effect of curvature in understanding
  the directionality of stripe patterns on fishes.
  Turing-like patterns are observed in microorganisms with
  elongated structure like Radiolaria. Analysis similar to that 
 adapted here may be suitable to study and mimic the patterns on 
 such viruses. Moreover, since RD models are being used
 to understand the self-organization of molecules on biological membranes, where toroidal
  and ellipsoid shapes are common, the above analysis thus may be useful
 to understand the curvature effects on self-organization in such membranes. It would
 be also interesting to extend our analysis to other 
 geometries, such as hyperboloid surface which seems to
 play an important role in biological materials~\cite{hyperbola1}.
 \section{ACKNOWLEDGEMENT}
\paragraph*{}
We acknowledge Sreedhar Dutta for suggesting the problem, and for useful discussions. We also
thank him for various helpful comments during the preparation of the manuscript. 

\appendix
\section{Geometric quantities on a torus and on an ellipsoid}
The torus can be parametrized as
\begin{equation*}
	X(\theta,\phi) = 
		\begin{pmatrix} 
			(R+r\cos \theta)\cos \phi \\ 
			(R+ r\cos \theta)\sin \phi \\ 
			r\sin \theta
		\end{pmatrix},
		\label{eq:3app}
\end{equation*}
and using the above parametrization one can read intrinsic and extrinsic quantities
related to curvature as
\begin{eqnarray*}
 g_{\theta\theta}=r^{2},\;g_{\phi\phi}=(R+r\cos\theta)^{2},\;g_{\theta \phi}=g_{\phi \theta}=0,\\
 \kappa_{\theta\theta}=-r,\;\kappa_{\phi\phi}=-(R+r\cos\theta)\cos\theta,
 \kappa_{\theta \phi}=\kappa_{\phi \theta}=0,
 \label{eq:12}
\end{eqnarray*}
and then using the intrinsic and extrinsic quantities, the gauss curvature $K$
and mean curvature $H$ on a torus can be read as
\begin{eqnarray*}
 K = \frac{\cos\theta}{r(R+r\cos\theta)},\\
 H=\frac{-(2r\cos\theta+R)}{2r(R+r\cos\theta)}.
 \label{eq:13}
\end{eqnarray*}
The ellipsoid can be parametrized as
\begin{equation*}
	X(\theta,\phi) = 
		\begin{pmatrix} 
			a\sin\theta\cos\phi \\ 
			a\sin\theta \sin\phi \\ 
			b \cos\theta
		\end{pmatrix},
		\label{eq:10:app}
\end{equation*}
and using the above parametrization we read intrinsic and extrinsic quantities
related to curvature as
\begin{eqnarray*}
 g_{\theta\theta}&=&a^{2}(\cos^{2}\theta+\frac{b^{2}}{a^{2}}\sin^{2}\theta)
 ,\;g_{\phi\phi}=a^{2}\sin^{2}\theta,\;g_{\theta \phi}=g_{\phi \theta}=0,\\
 \kappa_{\theta\theta}&=&\frac{b}{(\cos^{2}\theta+\frac{b^{2}}{a^{2}}\sin^{2}\theta)^{1/2}},\;
 \kappa_{\phi\phi}=\frac{b\sin^{2}\theta}{(\cos^{2}\theta+\frac{b^{2}}{a^{2}}\sin^{2}\theta)^{1/2}},\\
 \kappa_{\theta \phi}&=&\kappa_{\phi \theta}=0,
 \label{eq:14}
\end{eqnarray*}
 and then gauss and mean curvature on an ellipsoid is given by
\begin{eqnarray*}
 K = \frac{b^{2}}{a^{4}(\cos^{2}\theta+\frac{b^{2}}{a^{2}}\sin^{2}\theta)^{2}},\\
 H=b\frac{1+(\cos^{2}\theta+\frac{b^{2}}{a^{2}}\sin^{2}\theta)}{2 a^{2}
 (\cos^{2}\theta+\frac{b^{2}}{a^{2}}\sin^{2}\theta)^{3/2}}.
 \label{eq:15}
\end{eqnarray*}

 \bibliography{ref}
 
\end{document}